# New View of Seismic Risk Assessment and Reduction of Accumulated Tectonic Stress Based on Earthquake Prediction

**Manana Kachakhidze[1,2], Nino Kachakhidze-Murphy[2], Victor Alania[3], Onise Enukidze[3]**

[1]LEPL Institute "OPTIKA", Tbilisi, Georgia
[2]Georgian Technical University, Tbilisi, Georgia
[3]Ivane Javakhishvili Tbilisi State University, M. Nodia Institute of Geophysics, Tbilisi, Georgia

**Abstract**

A large earthquake is still considered to be a natural catastrophic event that not only causes immeasurably great material damage to seismically active regions and countries but also takes the lives of thousands of people. That is why the assessment of seismic risk is one of the most important issues, and research in this area is now in one of the first places in all countries. However, despite the high level of research, the practical results of the seismic risk do not provide great protection guarantees from disaster. Based on recent studies of earthquake prediction possibility, the offered article presents a completely different view on the possibilities of seismic risk assessing and reducing the accumulated tectonic stress in the large earthquake focus.

**Discussion**

Since, on the base of modern satellite and terrestrial observations, it has become possible to create a model of the generation of electromagnetic emissions detected prior to earthquakes and accordingly, analytically to connect the frequency of these radiations with the length of the major fault arising in the focus of a future earthquake, the possibility of short-term and operative prediction of an earthquake has become completely real (Kachakhidze et al., 2015).

Studies have unequivocally confirmed that the electromagnetic radiation before an earthquake is unambiguously the only field, that, with satisfactory accuracy, describes the avalanche-unstable process of fault formation during the earthquake preparation period which implies the different sizes fractures closing and opening process in the seismogenic zone and their ordering in a certain way, which ultimately leads to the formation of the main fault (Kachakhidze et al., 2015; Kachakhidze et al., 2019).

This means that the electromagnetic radiation, existing prior to the earthquake, even in the early stages of earthquake preparation, gives us information about any changes in the length of the major fault, or these are change in the magnitude of a future earthquake. Of course, this field, in addition to the magnitude, allows to determine the epicenter of the future earthquake and the time of the occurring of an expected earthquake, and therefore it should be considered as a precursor to the earthquake (Kachakhidze et al., 2019). Thus, the solution to two problems is on the agenda: an incoming large earthquake risk assessment and reduction of accumulated tectonic stress.

Given the modern advances in geodetic, geotectonic, and earthquake surveys, some changes need to be made in their solution capabilities, and the research phases should therefore be as follows:

### § 1. Seismic risk assessment

1.1. To detect geotectonic anomalies in seismically active regions (countries), it is necessary continuously to conduct geodetic surveys using GPS and GNSS methods, as well as quantitatively to estimate the earthquake preparation area. For example, a geotectonic anomaly identified by studies near Tbilisi has the following appearance (Fig. 1) (Sokhadze

et al., 2018):

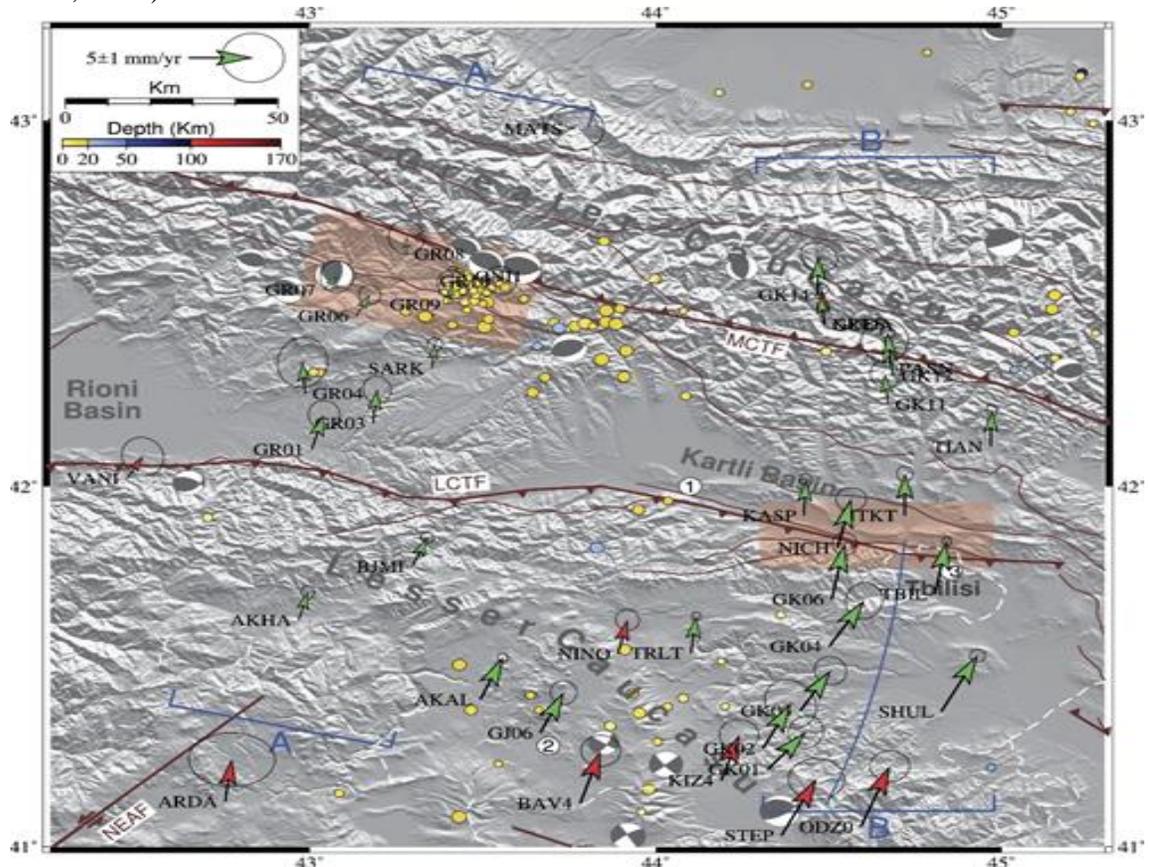

**Figure 1.** Map of GPS velocity, faults and focal mechanisms

1.2. The magnitude of the expected earthquake should be calculated approximately according to the estimated area by P. Dobrovolsky's (Dobrovolsky et al., (1979) formula (1):

$$R = 10^{0.43 M} \quad (1)$$

where R is the strain radius in km and M is earthquake magnitude

1.3. In case of an earthquake with an expected magnitude of M≥3, the earthquake prediction studying team should start working in the area and conduct surveys based on the VLF / LF EM emissions existing prior to the earthquake; These surveys should continuously monitor the magnitude of the expected earthquake, determine the coordinates of the future earthquake epicenter, estimate the time of the earthquake occurring, and the probable direction of the main fault in the earthquake focus.

1.4  The geometry of the active fault in the foci should be studied and the the displacement direction along fault should be specified. It is well-established, that surface geological records and deep drilling data are not sufficient for producing a true deep structure nature. Seismic profiles are essential in the introducing of a complex picture (Shaw et al., 2006). In the recent years integrated geological-geophysical surveys have been undertaken for Los-Angeles (USA) and Tokyo (Japan) megapolises where on the ground of seismic profiles interpretations so called blind faults, thrusts that are not exposed at the surface have been identified at depth. It was these hidden faults that were linked to the strong earthquakes of Los Angeles and Tokyo (Shaw et al.,, 1999; Sato et al., 2005).

   The success of these studies is largely determined by the application of fault-related folding theory (Shaw et al., 2006; Suppe et al., 1990). One of the components of this theory are fault-related folds (fault-bend and fault-propagation folds), the kinematics of which are mainly related to the geometry of the horizontal and ramp segments on which the displacements take place;

1.5 The crustal-scale structural cross-section of the area around the earthquake foci should be constructed using geological and geophysical methods of survey and taking into account the geometry and kinematics of the defined by above mentioned studies fault (or faults) in the earthquake focus.

For example, in the case of Georgia, for the construction of a structural model (or structural cross-section) 1:100,000 scale geological map, seismic profiles and tomographic data of regional earthquakes (Zabelina et al., 2016) will be used. For the construction of crustal-scale cross-section new numerical models of the Alpine-type orogens (Erdos et al., 2014; Grool et al., 2019) will be considered. The obtained structural model will make it possible to assess the active tectonic processes in the continental crust of the Lesser Caucasus. Unlike diagram geological cross-sections, structural cross-sections clearly reflect model deep structure, fold geometry and thrust trajectory.

The example provided in Figure 2 shows the lithospheric-scale structure of the Lesser Caucasus orogen.

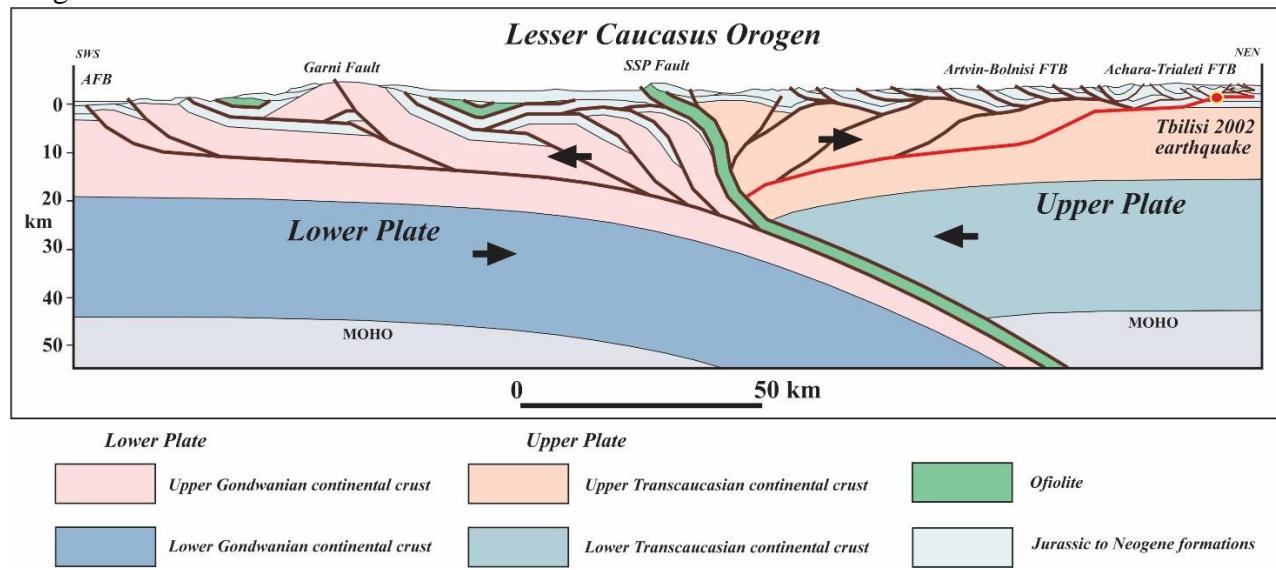

**Figure 2.** Lithospheric-scale structural cross-section across central Lesser Caucasus orogen (adapted from Alania et al., 2019 and Sosson et al., 2016).

For example, Figure 2 shows an active fault with a stepwise traectory. According to V. Alania (Alania et al., (2017) the upper segment of this fault should be connected to the Tbilisi 2002-04-25 17:41:21 (UTC) 41.77 ° N 44.86 ° E Ms = 4.3 (NEIC) earthquake.

1.6. Based on the studies, determine the preferred direction of seismic energy radiation from the incoming earthquake focus.

1.7. Seismic risk assessment should be specified based on the results obtained from the above-described studies.

The proposed method is completely different from the existing seismic risk assessment methodologies, as the existing methodology is based on retrospective data, but the work scheme presented by us is based on the possibility of short-term prediction of an incoming earthquake proceeds from the fact that we already can determine the epicentral area (and magnitude) of an incoming earthquake at least several dozen days before the earthquake It is obvious that seismic risk assessment method, proposed by us, will use today's existing risk assessment method, as a baseline, of course. But now we have the opportunity to correct and specify it for the concrete incoming earthquake, taking into account the specific magnitude, the time of the earthquake occurring, the coordinates of the earthquake focus, and the given specific geological structure. In addition, besides the other parameters required for a seismic risk assessment that the current methodology incorporates, the proposed study will allow us to determine, preferential direction of seismic energy radiation from an incoming earthquake focus.

**§ 2. Reduction of accumulated tectonic stress**

At the end of the last century, well-known Russian scientist, corresponding member of the Russian Academy, doctor of physics and mathematics Sciences Aleksey V. Nikolaev published an article "On the possibility of artificial removal of tectonic stress using seismic and electrical influences", where he notes that "one of the most important achievements of seismology of the last decade is associated with the study of the initiation of earthquakes" (Nikolaev, 1999). The work considered both "Seismicity initiated by natural processes "(Earth tide, changes in the speed of Earth's rotation, weather phenomena, large earthquakes), also "Seismicity initiated by technogeneous impacts" (Construction of huge dams and reservoirs, oil and gas production, underground nuclear tests, impact on the earth's crust with powerful electrical impulses, other technogeneous impacts) (Mirzoev et al., 2009; Nikolaev, 1995 (1), 1995 (2), 1997, Nikolaev et. Al., 1995, Gor'kovaty et al.,1994, Gupta, 1993, Sitinsky, 1989, Tarasov 1997; Tarasov et al., 1999).

The seismic reaction of the environment to the combined impact of several initiating natural and technogeneous (man-made) processes deserves special attention.

Some influences affect the focuses of that earthquakes, which would be inevitably occurred even without external intervention, but at the same time, the process of their preparation would take a little longer, and the magnitude would be somewhat greater". They have an effect on seismicity, reduce seismic hazard and change the ratio of the energy of large and weak earthquakes in favor of weak ones. But there are influences that cause earthquakes which would never have occurred, or would have occurred not very soon: these are the excavation of minerals, the construction of huge reservoirs, the injection of liquid industrial waste into the ground. These impacts increase the seismic hazard and, due to the expansion of the scale of engineering activities, industrial and civil construction, rapidly increase the seismic risk. Thanks to the exclusively high sensitivity of seismic processes, the question arises about the possible application of artificial influences on the earth's crust for the purpose of artificially guided discharge of tectonic stresses, the initiation of destructive earthquakes in predetermined intervals of time to reduce seismic hazard and risk. Reducing the seismic hazard should be carried out not only by reducing the magnitude of destructive earthquakes using special impacts on the earth's crust but also by reducing the hazards of destructive technogeneous earthquakes occurring. This can be achieved by choosing the optimal modes of technogenic impacts, coordinated as far as possible with natural impacts. Controlling the process of discharging tectonic stresses, artificially initiating the growth of weak and moderate seismicity, accelerating the preparation process of a strong earthquake, and as a consequence, reducing its magnitude should be considered possible (Nikolaev, 1999).

It is noteworthy that the author pays special attention to the powerful electrical impulses among the various methods of impact on the Earth's crust. The author notes that the response of seismicity to powerful electrical impulses is the most surprising result obtained in recent years. The gist is that the initiating electric field is very weak.

It was found that a powerful dipole source (hydromagnetic generator) with spread of electrodes of 1.5 km, emitting a 5-second pulse into the ground, the total energy of which is 10 MJ, noticeably excites seismicity in a radius of up to 50 km (region of Central Tajikistan). The number of earthquakes that occurred before and after running of electric signals were 15 and 28, respectively. A much more powerful source (100 MJ) is installed and operates in Kyrgyzstan. Its initiating effect on the seismicity of weak and moderate earthquakes (M = 1.5 ... 4) is clearly manifested at a distance of up to 500 km. The calculation of the additional seismic energy released as a result of the initiating action of electrical impulses is 5-6 orders of magnitude higher than the energy acting on the focal zones of earthquakes (Tarasov, et al., 1999).

The seismic response to electrical impact is observed during the first 2-3 weeks after the "electroshock", while mostly shallow (focal depth up to 5 km) earthquakes are initiated. Low-magnitude earthquakes (up to 1.5) and earthquakes with a magnitude of more than 3 respond most strongly to electrical exposure (Tarasov, et al., 1999).

Artificial discharge of tectonic energy can be done in two different ways. The first is the initiation of discharge in large areas by regular "processing" of these areas with powerful electrical and seismic sources, underground nuclear or chemical explosions. The second method is a purposeful impact on the focus of the incoming large earthquake. In addition, powerful pulsed electrical sources can be replaced by sources of relatively low power, but long-acting. However, the author had been noted that despite the fact that much attention has been paid to the study of induced seismicity in recent years, this problem remains poorly understood. To achieve this goal, it was necessary to find a place where nature has already prepared a catastrophic earthquake. In other words, it is necessary to solve the problem of forecasting an earthquake (Nikolaev, 1999).

To possess forecasting methods means to be able to predict the impact of natural and technogeneous influences on the development of an earthquake focus. This also means that the forecast of earthquakes, in combination with a controlled technogeneous impact, will allow to influence the process of focus preparation, stimulate the discharge of tectonic stresses, reduce the magnitude of destructive earthquakes and regulate the time of their occurrence. With approaching the moment of occurrence of an earthquake, the zone of focus becomes more and more sensitive to external influences. Therefore, the problem of intentional initiation of earthquakes is very closely related to the problem of earthquake prediction: first, it is necessary to identify the prepared focus, and then with the help of the impact it is possible significantly - up to several days (instead of weeks and months) – to reduce the time of occurrence of an earthquake (Nikolaev, 1999).

But today, based on our research (Kachakhidze et al., 2015; Kachakhidze et al., 2019) and by using special, reasonably organized impacts on the earthquake focus, it is possible to carry out a controlled discharge of tectonic energy, reduce seismic hazard (Mirzoev et al., 2009; Nikolaev, 1995 (1), 1995 (2), 1997, Nikolaev et. Al., 1995, Gor'kovaty et al.,1994, Gupta, 1993, Sitinsky, 1989, Tarasov 1997; Tarasov et al., 1999).

These studies are important because when studying the territory based on VLF / LF electromagnetic radiation, approximately 50 days before the earthquake, the magnitude of the earthquake is already approximately known. In addition, on the radiation channel of a given frequency is clearly shown the moment of onset and the developing character of the avalanche – unstable process of the earthquake preparation (Kachakhidze et. Al., 2019), which allows to select simple, easy implemented, environmentally safe, inexpensive method and handy time of impact for reducing of tectonic tension. Here we mean the possibility when the discharged method will affect not the entire area containing the tectonic anomaly, but a specific "live"fault.That is why we let ourselves to return to this vital issue again.By our view, in order to prevent large earthquakes, the time has come, when the formulation of the problem of controlled discharge of accumulated tectonic stress in the earth's crust, becomes actual.

Of course, working in individual countries, due to the whole region geological structure and the quality of the neighboring countries urban development, we should be very careful. For example, in (Fig. 2) is shown active fault, that caused the 2002 Tbilisi earthquake and besides Georgia, also has been extending over the territory of Armenia.

To avoid unforeseen difficulties, it is best to pre-detect a tectonically anomalous area in each seismically active region and, given the ability to predict earthquakes, exactly choose the time of safe impact to the earth's crust, for instance, with powerful electrical impulses to reduce the concrete tectonic stress. Naturally in this case, the choice of the moments of impacts should be carried out taking into account the influences of natural processes too. Although the scientific results obtained by Russian scientists still more require the continuation of deep and detailed research, in addition, it needs to be refined for each region (country) and for a specific large earthquake, the contours of the possibilities of reduction of accumulated tectonic stress are clearly outlined.

Today, at the modern stage of science development, the task of gradually easing the energy accumulated in the focus of a large earthquake for the reduction of accumulated tectonic stress is completely real and the major contribution in the development of research in this direction, no doubt, belongs to the member-correspondent of the Russian Academy, Dr. of Phys.-Math. Sci. A. V. Nikolaev and those scientists who have been working on this problem for many years. Thus, in our view, due to the possibility of short-term earthquake prediction, research works for the reduction of accumulated tectonic stress really may already be on the agenda.

Considering the problem of artificial discharge of tectonic stress the authors of the article unequivocally take into account Convention (ENMOD) prohibiting the military or other hostile use of environmental modification techniques causing earthquakes and tsunamis and having widespread, long-lasting, or severe effects (Convention, 1977).

As for the work stages that must be implemented to impact on the earth's crust with powerful electrical impulses, are the followings:

2.1. The items 1.1; 1.2; 1.3; 1.4 of §1 (Earthquake risk assessment), must be carried out;

2.2. Based on noted studies, the direction and form of the fault in the focus of the expected earthquake must be defined.

2.3. To identify possible initiating natural and technogenic processes affecting the focus.

2.4. To calculate the value and direction of the startup impulse for each impending earthquake. It must be considered the direction, shape of the fault existing in the earthquake focus and other parameters, which necessity will be revealed in future surveys;

2.5. Only after implementation of these tasks it is possible that the earth's crust to be impacted by using powerful electrical impulses, which will allow us gradually and safely to discharge the tectonic energy accumulated in the focus of an incoming large earthquake.

It should be emphasized that one of the crucial and necessary point in this work is item 1.3 of §1, which provides for studies based on VLF / LF EM emissions.

**Conclusions:**

The ability to predict large inland earthquakes, based on the researches of EM emissions existing prior to earthquakes, has radically changed the ways of solving the problems of earthquakes. It turned out that it is possible:
1. Among other parameters, it should be estimated the amount of possible radiated energy and radiated energy angle in the focus of a specific impending earthquake;
2. To assess the seismic risk for a concrete, already prepared impending earthquake;

3. In tectonically hazardous areas, where the large earthquake preparing process will be proved by studies, maximal reduce the tectonic tension artificially by the impact with powerful electrical impulses.